\documentclass[twocolumn,5p,number,letterpaper]{elsarticle}%
\usepackage{amsmath}
\usepackage{epsfig}
\usepackage{natbib}
\usepackage{accents}
\usepackage{amsfonts}
\usepackage{amssymb}
\usepackage{graphicx}%
\providecommand{\U}[1]{\protect\rule{.1in}{.1in}}
\providecommand{\U}[1]{\protect\rule{.1in}{.1in}}

\journal{Solid State Communications}
\begin{document}
%
\begin{frontmatter}
\title{Thermopower and Thermally Induced Domain Wall Motion in (Ga,Mn)As}
\author{Kjetil M. D. Hals\fnref{label1}}
\author{ Arne Brataas \fnref{label1}}
\author{Gerrit E. W. Bauer\fnref{label2}}
\address
[label1]{ Department of Physics, Norwegian University of Science and Technology, NO-7491 Trondheim, Norway}
\address
[label2]{Kavli Institute of NanoScience, Delft University of Technology,
Lorentzweg 1, 2628 CJ Delft, The Netherlands}
\begin{abstract}
We study two reciprocal thermal effects in the ferromagnetic
semiconductor (Ga,Mn)As by scattering theory: domain wall motion induced by a temperature gradient
as well as heat currents pumped by a moving domain wall. The effective out-of-plane thermal spin
transfer torque parameter $P_Q \beta_{Q}$, which governs the coupling between heat
currents and a magnetic texture, is found to be of the order of unity. Unpinned domain walls are predicted to move at speed 10 m/s in temperature gradients of the order 10 ${\rm K/ \mu m}$. The cooling power of a moving domain wall only compensates the heating due to friction losses at ultra-low domain wall velocities of about 0.07 m/s. The Seebeck coefficient is found to be of the order 100-500 ${\rm \mu V / K}$ at T=10 K, in good agreement with recent experiment.
\end{abstract}
\begin{keyword}
A. Ferromagnets; D. thermoelectrics; D. domain wall;  D. spin caloritronics
\PACS\ 72.15.Jf, 75.30.Sg, 75.78.-n, 75.50.Pp

\end{keyword}
\end{frontmatter}%

\begin{figure}[h]
\centering
\includegraphics[scale=0.7]{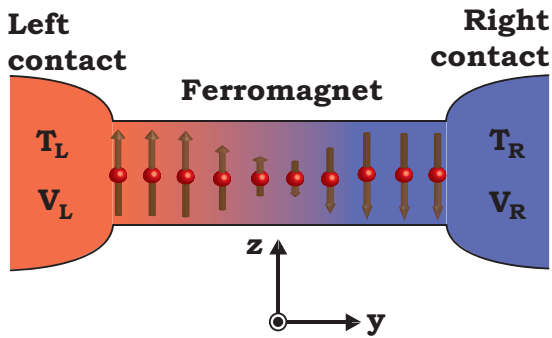}\caption{(Color online)\textbf{ }A
ferromagnetic wire containing a domain wall in which the magnetic texture
rotates in the transverse xz-plane (Bloch wall)~\cite{comment3}.}%
\label{DWsystem}%
\end{figure}

A magnetic domain wall is a region with a gradual reorientation of the local
magnetic moments between magnetic domains with different magnetization
directions (Fig.~\ref{DWsystem}). The position of a domain wall in a magnetic
wire can be manipulated by an external magnetic field or an electric
current~\cite{Tserkovnyak:jmmm08,Tatara:PR08}. The latter mechanism makes
electronic shift registers possible in which the magnetic domains separate
different bits which are collectively moved by a current (\textquotedblleft
racetrack memory\textquotedblright)~\cite{Parkin:Science08}.

When a current, which in a metallic ferromagnet is spin polarized, traverses a
magnetic domain wall, it exerts a torque on the domain wall, which can be
decomposed into an in-plane and an out-of-plane
component~\cite{Tserkovnyak:jmmm08}. In the adiabatic limit, the in-plane
torque component can be understood easily in terms of conservation of angular
momentum. The out-of-plane component is parameterized by a so-called $\beta
$-factor~\cite{Zhang:prl04}. Together with the Gilbert damping parameter
$\alpha$, which describes magnetic friction processes, it determines the
current-driven domain wall mobility: an increasing $\beta$-factor increases
the mobility, while it decreases with increasing $\alpha$%
~\cite{Tserkovnyak:jmmm08}.

The $\beta$-factor associated with a voltage gradient is denoted $\beta_{c}$.
In transition metal ferromagnets, $\beta_{c}$ and $\alpha$ are found to be of
the same order of magnitude ($\sim10^{-3}-10^{-2}$%
)~\cite{Kohno:jpsj06,Tserkovnyak:jmmm08}. These parameters have recently also
been studied in ferromagnetic
semiconductors~\cite{Kiet:prl07,Hals:prl09,Garate:prb09}, in which the strong
spin-orbit coupling in the valence band is found to be responsible for a
drastically larger $\beta_{c}/\alpha\sim100$~\cite{Hals:prl09,Garate:prb09}.

In the last few years a new research field called Spin Caloritronics, 
loosely speaking the study of thermal effects in spintronics, has
emerged leading to renewed interest in domain wall motion induced by temperature gradients.
The effect was first studied a couple of decades ago~\cite{Jen_Berger}, and
has recently enjoyed a renaissance~\cite{Hatami:prl07,Saslow:prb07,Kovalev:cm09,Bauer:cm09,Kovalev:ssc}. Since currents
can be induced by both voltage and temperature gradients, there are two
distinct contributions to $\beta$. (1) $\beta_{c}$ linked to a voltage gradient,
 and (2) the recently predicted thermal $\beta_{Q}$ associated with a
temperature gradient~\cite{Kovalev:cm09,Bauer:cm09,Xia}. \ Analogous to electrically
induced domain wall motion, the ability to move the domain wall by a temperature gradient
is controlled by the $\beta_{Q}/\alpha$-ratio~\cite{Kovalev:cm09,Bauer:cm09}.
It can arise from ballistic non-adiabaticity
in narrow domain walls \cite{Xia}. Here we discuss the adiabatic and
dissipative contribution to $\beta_{Q}$ by microscopic scattering theory based
on the Luttinger Hamiltonian of the GaAs valence band.

In (Ga, Mn)As, strong spin-orbit interaction complicates the
definitions of spin-dependent conductivites and Seebeck coefficients.
Instead, we calculate the well-defined quantities $P_Q\beta_Q$ and $P_c\beta_c$ ($P_c$ and $P_Q$ defined below). The $P_Q\beta_{Q}$ ($P_c\beta_{c}$ ) parameter we calculate, describes the effective out-of-plane thermal (electric) spin transfer torque parameter experienced by a rigidly moving domain wall, i.e. a global parameter depending on the magnetization profile, in particular the domain wall length.  

(Ga,Mn)As has a large thermopower with Seebeck coefficients of the order
$S\sim100-300~\mu V/K$~\cite{Pu:prl08}. Here we calculate the thermopowers of 
different alloys and find good agreement with experiments, implying a strong particle-hole
asymmetry in the system. Combined with the large $P_c \beta_{c}$ values found
before, we also expect a high $P_Q \beta_{Q}$ in these materials. Here we present
the first study of domain wall motion induced by a temperature gradient in the
ferromagnetic semiconductor (Ga,Mn)As. We find a large thermally induced
domain wall response with $P_Q\beta_{Q}\sim1$, which is the same order of
magnitude as $P_c\beta_{c}$. This leads to the prediction that domain walls move
at a speed of $10~%
\operatorname{m}%
/%
\operatorname{s}%
$ with temperature gradients as low as $100~%
\operatorname{K}%
/%
\operatorname{\mu m}%
$, much smaller than what has been predicted for transition metals
\cite{Bauer:cm09}.

The paper is organized as follows: We first review the derivation of
$P_Q\beta_{c}$ and $P_c\beta_{Q}$ in terms of the scattering matrix following
Refs.~\cite{Hals:prl09,Bauer:cm09}. Then $P_Q\beta_{Q}$ is calculated numerically
for disordered (Ga,Mn)As, which enables us to predict the temperature
gradients needed to move domain walls. We also discuss the heat pumping by
domain wall systems \cite{Kovalev:cm09,Bauer:cm09}.

Magnetization dynamics is well described phenomenologically by the generalized
LLG equation~\cite{Tserkovnyak:jmmm08}
\begin{equation}
\mathbf{\dot{m}}=-\gamma\mathbf{m}\times\mathbf{H}_{\mathrm{eff}}%
+\alpha\mathbf{m}\times\mathbf{\dot{m}}+\boldsymbol{\tau}, \label{LLG}%
\end{equation}
where $\mathbf{m}=\mathbf{M}(\mathbf{r},t)/M_{\mathrm{s}}$ is the unit
direction vector of the magnetization $\mathbf{M}$, $\mathbf{H}_{\mathrm{eff}%
}=-\delta F[\mathbf{M}(\mathbf{r},t)]/\delta\mathbf{M}$ is the effective field
given by the free energy functional $F[\mathbf{M}(\mathbf{r},t)]$, $\gamma$ is
(minus) the gyromagnetic ratio, and $\alpha$ the Gilbert damping constant. In
conducting ferromagnets, the itinerant quasiparticles are strongly coupled to
the magnetization via the exchange interaction. Through this coupling a spin
current can induce a torque $\boldsymbol{\tau}$ on a magnetic texture. The
torque can be generated by a temperature gradient $\Delta T\equiv T_L - T_R$ ($\boldsymbol{\tau}_{Q}$) or
a voltage gradient $\Delta V\equiv V_L - V_R$ ($\boldsymbol{\tau}_{c}$). To lowest order in
magnetization gradients these two torques have the
form~\cite{Kovalev:cm09,Bauer:cm09}
\begin{align}
\boldsymbol{\tau}_{c}  &  =\frac{\hbar P_{c} G}{AeS_{0}}\Delta V \left(  1-\beta
_{c}\mathbf{m}\times\right)  \partial_{y}\mathbf{m},\label{STT_c}\\
\boldsymbol{\tau}_{Q}  &  =\frac{\hbar P_{Q} T S G}{AeS_{0}}%
\frac{\Delta T}{T}\left(  1-\beta_{Q}\mathbf{m}\times\right)  \partial_{y}\mathbf{m},
\label{STT_Q}%
\end{align}
where $S_{0}=M_{\mathrm{s}}/\gamma$ is the spin density along $-\mathbf{m}$,
$G$ is the conductance, $S=-e\mathcal{L}T\partial_{E}\ln G$ is the Seebeck
coefficient, $\mathcal{L}=(k_{B}\pi)^{2}/(3e^{2})$ is the Lorenz constant and
$\partial_{E}\equiv\partial/\partial E$ at the Fermi energy, $e$ the modulus
of the electron charge, $2 P_c \equiv (\sigma^{\uparrow} - \sigma^{\downarrow})/(\sigma^{\uparrow} + \sigma^{\downarrow})$
where $\sigma^{\uparrow(\downarrow)}$ are the spin-dependent conductivities, $P_{Q}\equiv P_c + P_s\left(1-4P_c^2 \right)$ ~\cite{Kovalev:cm09} 
with $2P_s\equiv (S^{\uparrow} - S^{\downarrow})/(S^{\uparrow} + S^{\downarrow})$ where $S^{\uparrow(\downarrow)}$ are the
spin-dependent Seebeck coefficients,
and $A$ is the cross section area of the conductor. The
first terms in Eqs.~\eqref{STT_c} and \eqref{STT_Q} are the in-plane torques,
which describe the angular momentum transfer on $\mathbf{m}$ by a spin current
following the magnetic texture adiabatically. The second terms, parameterized
by $\beta_{c}$ and $\beta_{Q},$ describe dissipative out-of-plane torques.
$\beta_{Q},\beta_{c}$ and $\alpha$ govern the domain wall response to an
applied temperature or voltage gradient.

We adopt a magnetic free energy functional for a wire along the $y$-axis
\begin{align}
F[\mathbf{M}]  &  =M_{s}\int\!\!d\mathbf{r}\biggl(\frac{K_{s}}{2}%
[(\nabla\theta)^{2}+\sin^{2}(\theta)(\nabla\phi)^{2}]+\nonumber\\
&  \frac{K_{\bot}}{2}\sin^{2}\theta\sin^{2}\phi-\frac{K_{z}}{2}\cos^{2}%
\theta-H_{\mathrm{ext}}\cos\theta\,\biggr), \label{MagnEnergy}%
\end{align}
where $K_{s}$ is the spin wave stiffness, $K_{\bot}$ and $K_{z}$ are
anisotropy constants, $H_{\mathrm{ext}}$ is an external magnetic field, and
$\theta$ and $\phi$ are the polar and azimuth angles describing the local
magnetization directions, respectively. A local minimum of this functional is
a Bloch wall~\cite{comment3} rotating in the (transverse) $x\mathrm{-}z$ plane
with local magnetization direction given by $\cos\theta=\tanh([y-r_{w}%
]/\lambda_{w})$, $\sin\theta=1/\cosh([y-r_{w}]/\lambda_{w})$, where $r_{w}$ is
the position of the wall and $\lambda_{w}$ the domain wall length. The
position can be manipulated either by a magnetic field or a spin current. For
low current densities and magnetic fields the wall moves rigidly with constant
tilt angle $\phi$ and domain wall length $\lambda_{w}=\sqrt{K_{s}%
/(K_{z}+K_{\bot}\sin^{2}(\phi))}$, while the domain wall position $r_{w}$ is
given by:
\begin{equation}
\frac{\alpha\dot{r}_{w}}{\lambda_{w}}=-\gamma H_{\mathrm{ext}}-\frac{\hbar
P_{c}\beta_{c}G}{AeS_{0}\lambda_{w}}\Delta V - \frac{\hbar P_{Q}TS\beta_{Q}G}%
{AeS_{0}\lambda_{w}}\frac{\Delta T}{T}. \label{rDotEq}%
\end{equation}
Eq.~\eqref{rDotEq} is obtained by substituting the Bloch wall ansatz with a
time-dependent tilt angle $\phi(t)$ into Eq.~\eqref{LLG}, and consider the
regime where $d\phi/dt=0$. The values of $\Delta T /T$, $\Delta V$ and $H_{\mathrm{ext}%
}$ at which this regime breaks down are known as the Walker thresholds, above
which also the parameters $\lambda_{w}$ and $\phi$ have to be treated as
dynamic variables. In this paper we will always assume $\Delta T /T$, $\Delta V$ and
$H_{\mathrm{ext}}$ to be below the threshold fields, so that $r_{w}(t)$ is the
only time-varying parameter. In (Ga,Mn)As the Walker ansatz breaks down for
velocities typically of the order 10 $%
\operatorname{m}%
/%
\operatorname{s}%
$~\cite{Dourlat:prb08}.

Eq.~\eqref{rDotEq} shows that the magnetic system responds to temperature and voltage
gradients. From Onsager's reciprocal relations~\cite{Onsager:pr31} we 
know that the reciprocal processes heat and charge pumping by a moving
domain wall also exist, and that the response coefficients have to obey
Onsager symmetry relations. The response coefficients for heat and charge
pumped by a moving domain wall thus gives us expressions for the parameters
$P_Q\beta_{Q}$ and $P_c\beta_{c}$. This strategy is used in
Refs.~\cite{Hals:prl09,Bauer:cm09} to derive scattering matrix expressions for
$P_c\beta_{c}$ and $P_Q\beta_{Q}$. The response coefficients $L_{Qw}$ ($L_{cw}$) and
$L_{wQ}$ ($L_{wc}$), describing heat (charge) current pumped by a moving
domain wall and domain wall motion induced by a temperature (voltage)
gradient, respectively, are found from the parametric pumping
formula~\cite{Brouwer:prb98,Hals:prl09,Bauer:cm09}:
\begin{align}
L_{cw}  &  =\frac{e}{\hbar}\frac{\Im m\left\{  \mathrm{Tr}\left[
\frac{\partial S}{\partial r_{w}}S^{\dagger}\hat{\tau}_{z}\right]  \right\}
}{\mathrm{Tr}\left\{  \frac{\partial S}{\partial r_{w}}\frac{\partial
S^{\dagger}}{\partial r_{w}}\right\}  }\,,\label{Eq:L_cw}\\
L_{Qw}  &  = -\frac{e^{2}\mathcal{L}T^{2}}{\hbar}\frac{\partial_{E}\left(  \Im
m\left\{  \mathrm{Tr}\left[  \frac{\partial S}{\partial r_{w}}S^{\dagger}%
\hat{\tau}_{z}\right]  \right\}  \right)  }{\mathrm{Tr}\left\{  \frac{\partial
S}{\partial r_{w}}\frac{\partial S^{\dagger}}{\partial r_{w}}\right\}  }\,.
\label{Eq:L_Qw}%
\end{align}
Here, $S$ is the scattering matrix, $\text{Tr}$ the trace over propagating
modes in the leads, and $\hat{\tau}_{z}=\left[  \,\mathbf{\hat{1}%
}\ 0;\,0\ -\mathbf{\hat{1}}\right]  $ a $2N\times2N$ block-diagonal matrix,
where $\mathbf{\hat{1}}$ is the $N\times N$ unit matrix and $N$ is number of
propagating modes at the Fermi energy in each lead, and we assume a mirror symmetric system.
Eq. (\ref{Eq:L_cw}) and (\ref{Eq:L_Qw}) also assume a sufficient long wire where back-action of the induced currents on the domain wall motion can be neglected~\cite{Bauer:cm09}.
Onsager's reciprocal theorem implies the symmetry relations $L_{Qw}[\mathbf{m}%
]=L_{wQ}[-\mathbf{m}]$ and $L_{cw}[\mathbf{m}]=L_{wc}[-\mathbf{m}]$.
Eq.~\eqref{Eq:L_cw} and \eqref{Eq:L_Qw} give the following expressions for
$P_c\beta_{c}$ and $P_Q\beta_{Q}$:
\begin{align}
P_c\beta_{c}  &  = -\frac{\lambda_{w}}{4} \frac{ \Im m\mathrm{Tr}\left[  \frac{\partial
S}{\partial r_{w}}S^{\dagger}\hat{\tau}_{z}\right] }{ \mathrm{Tr}\left[
\hat{t}^{\dagger}\hat{t}\right] } ,\label{BetaFactor1}\\
P_Q\beta_{Q}  & = -\frac{\lambda_{w}}{4} \frac{ \partial_E\Im m\mathrm{Tr}\left[  \frac{\partial
S}{\partial r_{w}}S^{\dagger}\hat{\tau}_{z}\right]}{  \partial_E\mathrm{Tr}\left[
\hat{t}^{\dagger}\hat{t}\right] }  , \label{BetaFactor2}%
\end{align}
where  $\hat{t}$ is the transmission matrix.

In addition to the heat current pumped between the left and right contacts by
the moving domain wall, heat is also generated by the
magnetization damping. For slow magnetization
dynamics, this energy dissipation rate can be obtained from the scattering
matrix as well~\cite{Brataas:prl08,Hals:prl09}:
\begin{align}
\alpha=\frac{\hbar\lambda_{w}}{8\pi AS_{0}}\mathrm{Tr}\left[  \frac{\partial
S}{\partial r_{w}}\frac{\partial S^{\dagger}}{\partial r_{w}}\right]  .
\label{alphaFactor}%
\end{align}
The damping is second-order in the magnetization dynamics and therefore not 
controlled by Onsager's theory.

We neither took into account inelastic electron scattering nor temperature
dependence of the material parameters.
This is believed to be a good approximation for the highly conducting 
transition metals, but might become significant in magnetic semicondutors,
but a more detailed study is beyond the scope of this manuscript.

In the following, we focus on the (Ga,Mn)As system, and calculate $P_Q\beta_{Q}$
using Eq.~\eqref{BetaFactor2} and $P_c\beta_{c}$ by the methods reported in
Ref.~\cite{Hals:prl09}. We investigate the reciprocal effects:
\textit{Heat pumping by a moving domain wall} as well as \textit{torque
induced by a temperature gradient}.

To model the band structure of (Ga,Mn)As we use the Hamiltonian~\cite{Modell}
\begin{equation}
H=H_{L}+~\mathbf{h(r)}\cdot\mathbf{J}+V(\mathbf{r}), \label{Hamiltonian}%
\end{equation}
where $H_{L}$ is the $4\times4$ Luttinger Hamiltonian for zincblende
semiconductors in the spherical approximation. $\mathbf{J}$ is a vector of
$4\!\times\!4$ spin matrices for $J\!\!=\!\!3/2$ spins. The $\mathbf{h}%
\cdot\mathbf{J}$ term is a mean field approximation of the exchange
interaction between the itinerant holes and the local magnetic moment of the
Mn dopants. The exchange field $\mathbf{h}$ is antiparallel to the
magnetization direction $\mathbf{m}$. $V(\mathbf{r})=\sum_{i}V_{i}%
\delta(\mathbf{r}-\mathbf{R}_{i})$ is the impurity scattering potential, where
$\mathbf{R}_{i}$ is the position of impurity $i$, and $V_{i}$ its 
scattering strength~\cite{comment2}. The $V_{i}$ are randomly and
uniformly distributed in the interval $[-V_{0}/2,V_{0}/2]$.

The response coefficients $L_{cw}$ and $L_{Qw}$ are calculated from the
scattering matrix expressions in Eq.~\eqref{Eq:L_cw} and Eq.~\eqref{Eq:L_Qw}.
The scattering matrix is calculated numerically using a stable transfer matrix
method~\cite{Usuki:prb95}. Disorder effects are included fully and
nonperturbatively by the ensemble average $\left\langle L_{ij}\right\rangle
=\sum_{n=1}^{N_{I}}\left(  L_{ij}\right)  _{n}/N_{I},$ where $N_{I}$ is number
of different impurity configurations. All coefficients are averaged until an
uncertainty $\delta\left\langle L_{ij}\right\rangle =\sqrt{\left(
\left\langle L_{ij}^{2}\right\rangle -\left\langle L_{ij}\right\rangle
^{2}\right)  /N_{I}}$ of less than $10\%$ is achieved.

The mean free path $l$ for impurity strength $V_{0}$ is calculated by fitting
the average transmission probability $T=\left\langle G\right\rangle /G_{sh}$
to $T(L_{y})=l/(l+L_{y})$ ~\cite{Datta:book}, where $G_{sh}$ is the Sharvin
conductance and $\left\langle G\right\rangle $ is the conductance for a system
of length $L_{y}$. 

We consider a discrete (Ga,Mn)As system with transverse dimensions $L_{x}%
=19$~$%
\operatorname{nm}%
$, $L_{z}=15$~$%
\operatorname{nm}%
$ and $L_{y}\in\lbrack100,350]$~$%
\operatorname{nm}%
$
connected to infinite ballistic leads. 
The lattice constant is $1$~$%
\operatorname{nm}%
$, much less than the Fermi wavelength $\lambda_{F}\sim10$~$%
\operatorname{nm}%
$. The Fermi energy is $0.077$~$%
\operatorname{eV}%
$ when measured from the lowest subband edge. The Luttinger parameters in
$H_{L}$ are $\gamma_{1}=7.0$ and $\gamma_{2}=2.5$, and $|\mathbf{h}|=0.032$~$%
\operatorname{eV}%
$~\cite{Modell}. $\lambda_{w}\in\left\{  10,20,40\right\}  $~$%
\operatorname{nm}%
$ are the wall widths.  To
estimate a typical saturation value of the magnetization we use $M_{s}%
=10|\gamma|\hbar x/a_{\mathrm{GaAs}}^{3}$~\cite{comment4}, with $x=0.05$ as
the doping level, and where $a_{\mathrm{GaAs}}$ is the lattice constant for GaAs.

Since we used the Sommerfeld approximation,  the scattering matrix expressions for the
Onsager coefficients in Eq.~\eqref{Eq:L_cw} and Eq.~\eqref{Eq:L_Qw} are valid
only for thermal energies that are small compared to the Fermi energy. For instance
$k_{B}T/E_{F}=0.01$, implies a temperature of the order $10~%
\operatorname{K}%
$ with the Fermi energy specified above.

\begin{figure}[h]
\centering
\includegraphics[scale=1.1]{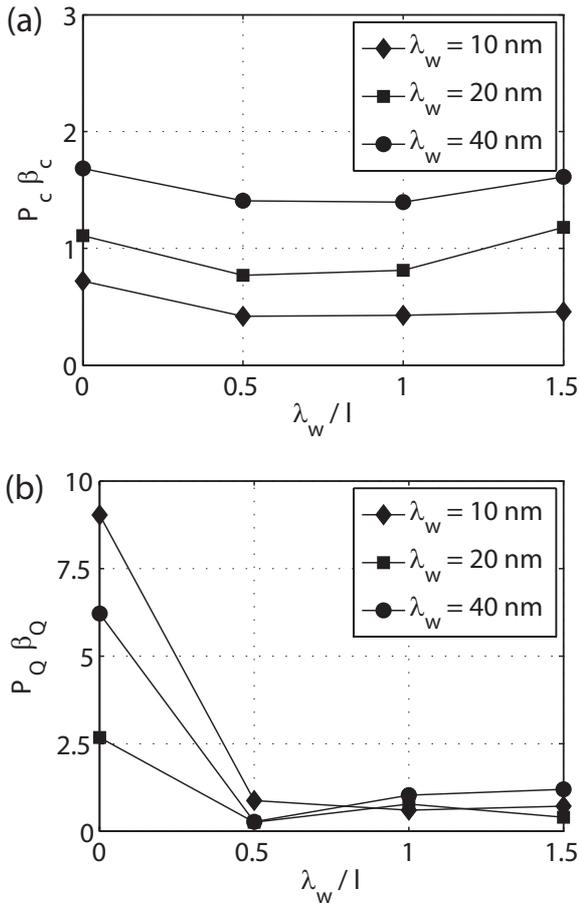}\caption{(a) $P_c\beta_{c}$ and (b)
$P_Q \beta_{Q}$ plotted as a function of domain wall length ($\lambda_{w}$) over
mean free path ($l$). The lines are guides to the eye.}%
\label{Fig1}%
\end{figure}

Fig.~\ref{Fig1} (a) and (b) show the out-of-plane torque parameters $P_c\beta
_{c}$ and $P_Q\beta_{Q}$ as a function of disorder. We observe that in the
diffuse regime $\left(  l\ll\lambda_{w}\right) $ $P_Q\beta_{Q}\sim1.0$ is of the
same order as $P_c\beta_{c}$ . As explained in Ref.~\cite{Hals:prl09}, the high
$P_c\beta_{c}$ of (Ga,Mn)As arises from large hole reflection at the domain wall
due to the intrinsic domain wall resistance caused by spin-orbit
coupling~\cite{Kiet:prl06}. This intrinsic resistance decreases with disorder,
reducing $P_c\beta_{c}$ with increasing disorder. For even smaller mean-free
paths, $P_c\beta_{c}$ saturates or increases slightly due to the increasing
spin-flip rate caused by the impurity scattering. In the ballistic limit
($l\gg\lambda_{w}$) $P_Q\beta_{Q}$ is nearly  one order of magnitude larger than
$P_c\beta_{c}$, but decreases rapidly when disorder is included, saturating in the
diffusive regime. Since $P_Q\beta_{Q}$ is a measure of the effect of particle-hole asymmetry on the
out-of-plane torque efficiency, this means that the particle-hole asymmetry of the torque efficiency is
stronger in the ballistic than in the diffuse regime.

In the diffuse regime we find Seebeck coefficients $S=100-500~%
\operatorname{\mu V}%
/%
\operatorname{K}%
$ ($S/T=10-50~%
\operatorname{\mu V}%
/%
\operatorname{K}%
^{2}$) at $T=10~%
\operatorname{K}%
$ for mean free path $l\sim7 - 27$~nm~\cite{Hals:Seebeck}, in good agreement with the experimental
values $S\sim100-300~%
\operatorname{\mu V}%
/%
\operatorname{K}%
$~\cite{Pu:prl08}. The Gilbert damping calculated from Eq.~\eqref{alphaFactor}
is of the order $10^{-3}-10^{-2}$~\cite{Hals:prl09}.

\begin{figure}[h]
\centering
\includegraphics[scale=1.1]{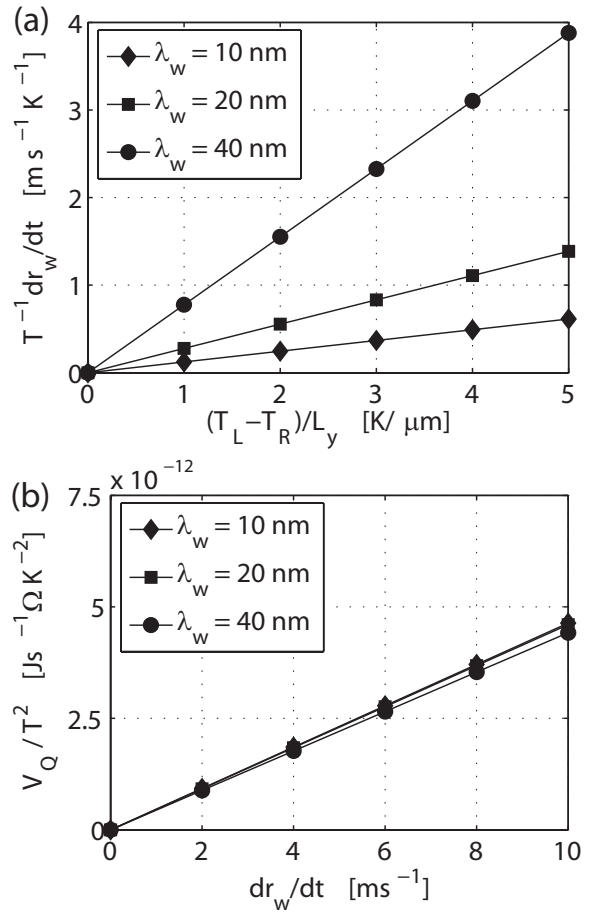}\caption{(a) Plot of domain wall velocity
divided by temperature induced by the temperature gradient $(T_{L}%
-T_{R})/L_{y}$. (b) The \textquotedblleft thermal motive
force\textquotedblright\ $V_{Q}=J_{Q}/G$ is displayed as a function of domain
wall velocity. In both plots the Seebeck coefficient is of the order
$S/T=10-50~\operatorname{\mu V}/\operatorname{K}^{2}$ and the conductivity
$\sigma= 270-630~\operatorname{\Omega}^{-1}\operatorname{cm}^{-1}$.}%
\label{Fig2}%
\end{figure}

The high $P_Q\beta_{Q}$ combined with a large Seebeck coefficient implies a
strong coupling between thermally induced currents and domain wall motion in (Ga,Mn)As.
This increases the ability to manipulate the domain wall position with a temperature gradient,
and oppositely, to transfer heat between the left and right contacts
by moving the domain wall. These two effects are illustrated in
Fig.~\ref{Fig2}, where we plot the reciprocal thermal effects: Domain wall
motion induced by a temperature gradient and heat current pumped by a moving
domain wall. Fig.~\ref{Fig2}~(a) shows the domain wall velocity divided by $T$
as a function of temperature gradient, while Fig.~\ref{Fig2}~(b) shows the
\textquotedblleft thermal motive force\textquotedblright\ $V_{Q}=J_{Q}/G,$ a
system-size independent quantity determining how efficient the domain wall
system pumps a heat current in the absence of a temperature or voltage bias,
divided by $T^{2}$ as a function of domain wall velocity.

To get a better feeling of the importance of these two reciprocal processes,
let us consider a wire at temperature $T=10~%
\operatorname{K}%
$ with $L_{y}=60~%
\operatorname{\mu m}%
$, and conductivity $\sigma\sim270~%
\operatorname{\Omega }%
^{-1}%
\operatorname{cm}%
^{-1}$, containing a Bloch wall with $\lambda_{w}=10~%
\operatorname{nm}%
$. We limit attention to the regime in which the domain wall moves rigidly,
\textit{i.e}. below the Walker thresholds. We see from Fig.~\ref{Fig2}~(a)
that a temperature gradient of around $10~%
\operatorname{K}%
/%
\operatorname{\mu m}%
$ is needed to drive the domain wall to speed $10$ m/s. According to
Eq.~\eqref{rDotEq}, an electric field-induced current density of the order
$10^{4}~%
\operatorname{A}%
/%
\operatorname{cm}%
^{2}$ will give an equivalent torque. However, in a real (Ga,Mn)As system, we
know that current densities of the order $10^{5}-10^{6}~%
\operatorname{A}%
/%
\operatorname{cm}%
^{2}$ are needed to achieve such an effect~\cite{Yamanouchi:nature04}. The disagreement is caused by reduction
of the domain wall mobility by extrinsic pinning effects~\cite{DWpinning}. Thus, taking into account this effect, one may expects that a temperature gradient of the order $100-1000~%
\operatorname{K}%
/%
\operatorname{\mu m}%
$ needed for driving the wall at speed $10~%
\operatorname{m}%
/%
\operatorname{s}%
$ is a more realistic estimate.

As mentioned above, there is also Joule
heat generation associated with the magnetization damping. Note that this
heating does not lead to a temperature difference between the left and right
contact for a mirror symmetric system since an equal amount of energy is
pumped into the right and left contact, but equally heats both two contacts.
Assuming a Gilbert damping of the order $\alpha=10^{-2}$ implies that the
domain wall dissipates a heat current density of $30$ $\mathrm{W/ m^{2}}$  due to magnetic
friction~\cite{Hals:prl09} into each reservoir, while the moving domain wall pumpes 
$0.2$ $\mathrm{W/ m^{2}}$ between the left and right contact. Thus, close to the Walker threshhold the effect of the
magnetic heat pump is rather small compared to the magnetic friction
process. 
The heat current density pumped by the moving domain wall ($j_{Q}$)
scales linearly with $\dot{r}_{w}$, while the energy current density
associated with magnetization damping ($j_{Q}^{\mathrm{mag}}$) scales
quadratically. For instance, for the system considered above $j_{Q} = 0.02\,
\dot{r}_{w}~\mathrm{s W/m^{3}}$ and $j_{Q}^{\mathrm{mag}} = 0.3\, \dot{r}%
_{w}^{2}~\mathrm{s^2 W/m^{4}}$. The domain wall system is therefore not an
efficient cooler except for very small domain wall velocities ($\dot{r}%
_{w}<0.07~\text{m/s}$), where the domain wall pumped energy will dominate over
the magnetic friction process. The situation is the same in transition metal ferromagnets, 
in which the out-of-plane torque parameter is of the same order as the Gilbert
damping~\cite{Tserkovnyak:jmmm08}. Insulating ferromagnets such as the Yttrium-Iron-Garnets have very
low Gilbert damping and are possibly better suited for cooling purposes.

In conclusion, we have studied thermally induced domain wall motion 
, and the reciprocal effect, a heat pump operated by moving the domain
wall, in the ferromagnetic semiconductor (Ga,Mn)As. The $P_Q\beta_{Q}$ parameter 
that governs both effects is found to be of the order unity in diffuse
systems. We estimate that a domain wall velocity of $10~%
\operatorname{m}%
/%
\operatorname{s}%
$ can be induced by a temperature gradient of the order $100-1000~%
\operatorname{K}%
/%
\operatorname{\mu m}%
$, while a domain wall pumps around $0.2~%
\operatorname{W}%
/%
\operatorname{m}%
^{2}$ when it moves at speed $10~%
\operatorname{m}%
/%
\operatorname{s}%
$. The energy dragged by the domain wall from one contact to the other is
relatively small compared to energy dissipation associated with magnetic
friction at these velocities, implying that (Ga,Mn)As domain wall systems are not
useful for cooling purposes.

We thank Anh Kiet Nguyen for developing the numerical transfer matrix code.
This work was supported in part by computing time through the Notur project
and EC Contract IST-033749 "DynaMax".


\begin{thebibliography}{99}
\bibitem {comment3}Our approach also holds for head-to-head (tail-to-tail)
domain walls.
\bibitem {comment2}In the discrete version of Eq.~\eqref{Hamiltonian} as used
in the numerical calculation we have one impurity at each lattice site.
\bibitem {comment4}The prefactor of 10 comes from 4 Ga atoms per unit cell
times spin $5/2$ per substitutional Mn, which are assumed to be fully
polarized. The reduction of the net magnetization due to the interstitial Mn
ions and $p$ holes are disregarded in our estimate, which is therefore a clear
overestimate. A reduced magnetization strongly increases the current-induced
effects as can be appreciated from the domain wall equation of motion
(\ref{rDotEq}).
\end{thebibliography}
\end{document}